\documentclass[ecrc,1p,eps]{elsarticle}

\usepackage{lineno,hyperref}
\usepackage{amssymb}
\usepackage{float}
\usepackage{pifont}
\usepackage{natbib}
\usepackage{tikz}
\usepackage{graphicx}
\usepackage{epstopdf}
\usepackage{url}
\usepackage{amsmath}
\usepackage{amsthm}
\usepackage[utf8]{inputenc}
\usepackage[T1]{fontenc}
\usepackage{lmodern}
\usepackage{ae}
\usepackage{enumerate}

\newtheorem{theorem}{\bf Theorem}[section]
 \newtheorem{proposit}[theorem]{\bf Proposition}
 \newtheorem{coro}[theorem]{\bf Corollary}
\newtheorem{lem}[theorem]{\bf Lemma}

\newtheorem{notat}{\bf Notation}
\newtheorem{remark}[theorem]{\bf Remark} 
\newtheorem{condit}{\bf Condition}
\newtheorem{fait}{\bf Claim}

\def\thm#1\par{\medskip\par\noindent\begin{theorem} \strut \sl #1 \end{theorem}\par}
\def\propo#1\par{\medskip\par\noindent\begin{proposit} \strut \sl #1 \end{proposit}
\par}
\def\cor#1\par{\medskip\par\noindent\begin{coro} \strut \sl #1 \end{coro}\par}
\def\lm#1\par{\medskip\par\noindent\begin{lem} \strut \sl #1 \end{lem}\par}
\def\defil#1\par{\medskip\par\noindent\begin{condit} \strut \sl #1 \end{condit}\par}
\def\fct#1\par{\medskip\par\noindent\begin{fait} \strut \sl #1 \end{fait} \cqfd\par}
\def\defi#1\par{\medskip\par\noindent{\begin{defin} \strut  \sl #1 \end{defin}}\par}
\def\nota#1\par{\par\noindent\begin{notat} \nopagebreak  \strut #1  \end{notat}}
\def\rem#1\par{\par\noindent\begin{remark} \nopagebreak \strut \rm #1   \end{remark}}
\def\ex#1\par{\par\noindent\begin{exemple} \nopagebreak \strut #1  \end{exemple}}
\newdefinition{rmk}{Remark}
\newdefinition{expl}{Example}
\newproof{pf}{Proof}

\def\N{\mbox{I\hspace{-.15em}N}}
\def\Z{\mbox{I\hspace{-.3em}Z}}

\def\Z{\mbox{l\hspace{-.47em}C}}

\def\cqfd{~~~~~~~~~~$\Box$}

\def\N{{\mathchoice {\hbox{$\sf\textstyle N\kern-0.4em N$}}
{\hbox{$\sf\textstyle I\kern-0.42em N$}}
{\hbox{$\sf\scriptstyle I\kern-0.2em N$}} 
{\hbox{$\sf\scriptscriptstyle I\kern-0.em N$}}}}
\def\Z{{\mathchoice {\hbox{$\sf\textstyle Z\kern-0.4em Z$}}
{\hbox{$\sf\textstyle Z\kern-0.4em Z$}}
{\hbox{$\sf\scriptstyle Z\kern-0.3em Z$}}
{\hbox{$\sf\scriptscriptstyle Z\kern-0.2em Z$}}}}
\theoremstyle{plain}

\begin{document}

\begin{frontmatter}
\title{Embedding a $\theta$-invariant code into a complete one}
\author{Jean N\'eraud}\corref{mycorrespondingauthor}
 \ead{jean.neraud@univ-rouen.fr, neraud.jean@gmail.com}
\cortext[mycorrespondingauthor]{Corresponding author: neraud.jean@gmail.com}
\author{Carla Selmi}
\ead{carla.selmi@univ-rouen.fr}
\address{ Laboratoire d'Informatique, de Traitemement de l'Information et des Syst\`emes (LITIS), Universit\'e de Rouen Normandie, UFR Sciences et Techniques,
Avenue de l'universit\'e, 76830 Saint Etienne du Rouvray, France}

\begin{abstract}
Let $A$ be an arbitrary alphabet and let $\theta$ be an (anti-)automorphism of $A^*$ (by definition, such a correspondence is determinated by a permutation of the alphabet). This paper deals with sets which are invariant under $\theta$ ($\theta$-invariant for short) that is, languages $L$ satisfying $\theta(L)\subseteq L$.
We establish an extension of the famous defect theorem. With regard to the so-called notion of completeness, we  provide a series of examples of finite complete $\theta$-invariant codes. 
Moreover, we establish a formula which allows to embed any  non-complete $\theta$-invariant code into a complete one. 
As a consequence, in the family of the so-called   thin $\theta$-invariant codes, maximality and completeness are two equivalent notions.
\end{abstract}


\end{frontmatter}
\section{Introduction}
\label{INTRO}
In the free monoid theory, during the last decade, research involving one-to-one {\it morphic} or {\it antimorphic} substitutions has played a particularly important part: this is due to the powerful applications of these objects, in particular in the framework of DNA-computing.
In the case of automorphisms or anti-automorphisms -for short we write \-{\it (anti-)automorphisms}- given an arbitrary {\it alphabet}, say $A$, any such  mapping is  completely determined by extending a unique  permutation of $A$ to $A^*$, the {\it free monoid} that  is generated by $A$.

In the special case of {\it involutive}  (anti-)automorphisms, lots of successful investigations have been done 
for extending most of the now classical  combinatorial properties on  words.
The topics of the so-called {\it pseudo-palindromes} \cite{dD06}, that of {\it $\theta$-episturmian} words \cite{BdZD11}, and the one of {\it pseudo-repetitions} \cite{ARVS14,GMRMNT13} have been particularly involved.
The framework of  some peculiar families of {\it variable-length codes} \cite{KM06} and that of {\it equations in words} \cite{CCKS11,DFMN17,KM08,MMNS14} have been concerned.
Generalizations of the famous theorem of Fine and Wilf (\cite{FW65},\cite[Proposition 1.3.5]{Lo1983}) were also established \cite{CKS10,MMN11}.

Equations in words are also the starting point of the study in the present paper, which consists in some full version of \cite{NS17}.
Let $A$ be an arbitrary alphabet and let $\theta$ be an  (anti-)automorphism of $A^*$; we adopt the point of view from \cite[Ch. 9]{Lo1983}, by considering a finite collection of unknown words, say $Z$.
We assume that a (minimum)  positive integer $k$ (i.e. the so-called {\it order} of $\theta$) exists such that $\theta^k=id_{A^*}$.
This condition  is particularly satisfied by every (anti-)automorphism whenever $A$ is finite.
In view of making the present forward more easily readable,  in the first instance let us take $\theta$  as an involutive (anti-)automorphism (that is, $\theta^2=id_{A^*}$).
We assign that the words in $Z$ and their images by $\theta$ to satisfy a given equation, and we ask for the computation of a finite set of words, 
say $Y$, such that  all the words of $Z$ can be expressed as a concatenation of words in $Y$. 
Actually, such a question  appears more complex than in the classical configuration, where $\theta$ does not interfer:
in this classical case, according to the famous defect theorem \cite[Theorem 1.2.5]{Lo1983}, 
it is well known that at most $|Z|-1$ words  allow to compute the words in $Z$.  
At the contrary, due to the interference of  (anti-)automorphisms, in \cite{KM08},  examples where $|Y|=|Z|$ are provided by the authors.

Along the way, for solving our problem, applying the defect theorem  to  the set $X=Z\cup\theta(Z)$  might appear natural.   Such a methodology garantees 
the existence of a set $Y$, with $|Y|\le |X|-1$ and whose elements allow by concatenation  to rebuild all the words in $X$.
It is also well known that $Y$ can be chosen in such a way that only trivial equations may  hold among its elements: 
with the terminology of \cite{BPR10,Lo1983,Lo2002}, $Y$ is a {\it code}, or equivalently  $Y^*$, the submonoid that it generates, is {\it free}. 
Unfortunately, since both the words in $Z$ and $\theta(Z)$ are expressed  as concatenations of 
words in $Y$,  among the words of $Y\cup\theta(Y)$ non-trivial equations can still hold. In other words, by applying that methodology, 
the initial problem would be transferred among the words in $Y\cup\theta(Y)$.

An alternative methodology will consist in asking for codes $Y$ which are invariant under $\theta$ ($\theta$-{\it invariant} for short) that is, satisfying $\theta(Y)=Y$. 
Returning to the general case, where $\theta$ is an arbitrary (anti-)automorphism, this is equivalent to say that
the union of the sets $\theta^i(Y)$, for all $i\in {\mathbb Z}$,  is  $\theta$-invariant.
By the way, it is straightforward to show that the intersection of an arbitrary family of free $\theta$-invariant submonoids is itself a free 
$\theta$-invariant submonoid. In the present  paper we prove the following result:
\begin{flushleft}
{\bf Theorem 1.}
{\it Let  $\theta$ be an (anti-)automorphism of $A^*$  and let $X$ be a  finite $\theta$-invariant set.
If $X$  it  is not a code, then the smallest  $\theta$-invariant free submonoid of $A^*$ containing $X$ 
is generated by a $\theta$-invariant code $Y$, which furthermore satisfies  $|Y|\le |X|-1$.
}
\end{flushleft}
For illustrating this result in terms of equation, we refer to \cite{CCKS11,MMNS14}, where
the authors  considered generalizations of the famous  three unknown variables equation of Lyndon-Sh\"utzenberger  \cite[§     9.2]{Lo1983}. 
They proved that, an involutive (anti-)automorphism $\theta$ being fixed, given such an  equation with sufficiently long members, 
a word $t$ exists such that any 3-uple of ``solutions" can be expressed as a concatenation of words in $\{t\}\cup \{\theta(t)\}$.
With the notation of Theorem 1, the elements of the $\theta$-invariant set $X$ are $x,y,z,\theta(x),\theta(y),\theta(z)$ 
and those of $Y$ are $t$ and $\theta(t)$: we verify that, in every case $Y$ is a $\theta$-invariant code, furthermore we have $|Y|\le |X|-1$.\\

With regard to the theory of codes,  completeness is one of the most challenging notions: a subset $X$ of the free monoid $A^*$ is {\it complete} if any word is a factor of some word in $X^*$.
Maximality is another important notion: a code is {\it maximal} if it cannot be strictly included in some other code of $A^*$.  Actually, according to Zorn's Lemma, any code is included in  a maximal one 
moreover, a famous result due to Sch\"utzenberger states that, for the family of the so-called {\it thin} codes (which contains the regular codes), maximality and completeness are two equivalent notions \cite[Theorem 2.5.16]{BPR10}.  
From this point of view, in the second part of our study we are interested in  {\it complete} $\theta$-invariant codes.
It is natural to prealably examine the case of finite  codes. Clearly, the well-known complete {\it uniform codes} that is, the codes $A^n$ (with $n\ge 1$),
are invariant under every (anti-)automorphism.
Beside that, non-trivial  finite complete $\theta$-invariant codes exist:
for instance, take for $A$ the binary alphabet $\{a,b\}$, choose for $\theta$ the anti-automorphism that swaps the letters $a$ and $b$, and consider the complete code which was introduced in \cite{C72}: 
$$X=\{a^3,ab, a^2ba,a^2b^2, ba^2,baba,bab^2,b^2a,b^3\}.$$
It is straightforward to verify that $X$ is $\theta$-invariant.
In our paper, we provide some other examples: each of the  classes of bifix codes, prefix codes, and non-prefix non-suffix codes is concerned.

Despite that,  the question of describing a  general  structure for finite complete $\theta$-invariant codes remains largely open:
this is not surprizing since, with the exception of certain special families (e.g. \cite{DF89,DFR85,RSS89}), no general structure that could embrace finite complete codes is described in the literature. 

Another issue could consist in developing methods for embedding a code into a complete one.
However, in \cite{R75}, the author presents a class of codes that cannot be embedded into any finite complete one.
With regard to $\theta$-invariance, as far as we know, the question of embedding  finite codes into complete ones remains open.

Actually, in  \cite{R75},  the  question whether any finite code can be embedded into a regular one was  implicitely asked:
a positive answer was  brought in \cite{ER85}, where the authors provided a now classical  formula for embedding any regular code into a complete one.
In the present paper, we put a corresponding problem in the framework of $\theta$-invariant codes. 
Actually, by establishing the following result, we bring a positive answer:
\begin{flushleft}
{\bf Theorem 2.}
{\it 
Any non-complete $\theta$-invariant code  $X\subseteq A^*$,  can be embedded into a complete one.
Moreover, if $A$ is finite and $X$ regular, then $X$ can be embedded into a regular complete $\theta$-invariant code.\\
}
\end{flushleft}
As a consequence, we obtain the following result: it states that, in the framework of $\theta$-invariant codes, a property similar to a famous one due to Sch\"utzenberger  \cite[Theorem 2.5.16]{BPR10} holds:
\begin{flushleft}
{\bf Theorem 3.}
{\it Given a thin $\theta$-invariant code $X\subseteq A^*$,  the five following conditions are equivalent:

{\rm (i)} $X$ is complete.

{\rm (ii)} $X$ is a maximal code.

{\rm (iii)} $X$ is maximal in the family of the $\theta$-invariant codes.

{\rm (iv)} A positive Bernoulli distribution $\pi$ exists such that $\pi(X)=1$. 

{\rm (v)} For any positive Bernoulli distribution $\pi$, we have $\pi(X)=1$.
}
\end{flushleft}

We now describe the contents of our paper. 
Section 2 contains the preliminaries: the terminology of the free monoid is settled, 
and we recall some classical notions and results concerning  the codes.
The preceding  Theorem 1 is established in Section 3, 
where an original example of equation  is studied.
In Section 4, we present several examples of finite complete $\theta$-invariant codes.
The problem of embedding a finite $\theta$-invariant code into a complete one is also discussed:
this ensures a transition to the question of embedding a regular $\theta$-invariant code into a complete one.
This last question  
is studied in Section 5,  
where the preceding Theorem 2 and Theorem 3 are established.
\section{Preliminaries}
\label{preliminaries}
\subsection{Words and free monoid}
We adopt the notation of the free monoid theory.
In the whole paper, we consider an alphabet $A$, and  we denote by $A^*$ the free monoid that it generates. 
Given a word $w\in A^*$, we denote by $|w|$ its length, the empty word, which we denote by $\varepsilon$, being the word with length $0$.  
Given  a subset $X$ of $A^*$, we denote by $X^*$ the submonoid of $A^*$ that is generated by $X$, moreover we set $X^+=X^*\setminus \{\varepsilon\}$.

Let $x\in A^*$ and $w\in A^+$.  We say that $x$ is a {\it prefix} ({\it suffix}) of $w$ if a word $u$ exists such that $w=xu$ ($w=ux$).
Similarly, $x$ is a {\it factor} of $w$ if two words $u,v$ exist such that $w=uxv$. 
Given a non-empty set $X\subseteq A^*$, we denote by $P(X)$ ($S(X)$, $F(X)$) the set of the words that are prefix (suffix, factor) of some word in $X$. Clearly, we have $X\subseteq P(X)\subseteq F(X)$ ($X\subseteq S(X)\subseteq F(X)$). 
Given a pair of non-empty words $w,w'$, we say that it {\it overlaps} if  words $u,v$ exist such that $uw'=wv$ or $w'u=vw$,
with $1\le |u| \le |w|-1$ and  $1\le |v| \le |w'|-1$;  
otherwise, the pair is {\it overlapping-free} (in such a case, if $w=w'$, we simply say that $w$ is overlapping-free).

\subsection{Variable length codes} It is assumed that the reader has a fundamental understanding  with the main concepts of the theory of variable-length codes: we only recall some of the main definitions and we
suggest, if necessary,  that he (she) report to  \cite{BPR10}.
A subset $X$ of $A^*$ is a {\it variable-length code} (a {\it code} for short) if any equation among the words of $X$ is trivial 
that is, for any pair of sequences of words in $X$, say  $(x_i)_{1\le i\le n}$, $(y_j)_{1\le j\le p}$, the equation
$x_1\cdots x_n=y_1\cdots y_p$ implies  $n=p$ and $x_i=y_i$, for each integer $i\in [1,n]$.
By definition $X^*$ is a {\it free} submonoid of $A^*$.

In the present paper the so-called  {\it  prefix}, {\it suffix} and {\it bifix} codes play an noticeable part: a code $X\subseteq A^*$ is prefix (suffix) if $X\cap XA^+=\emptyset$ ($X\cap A^+X=\emptyset$).
A code is bifix if it is both prefix and suffix.

A code $X\subseteq A^*$  is {\it maximal} if it is not strictly included in some other code of $A^*$. 
Given a set $X\subseteq A^*$, it is {\it complete} if $A^*=F(X^*)$; $X$ is  {\it thin} if  $A^*\neq  F(X)$.  Regular codes are well known examples  of thin codes \cite[Proposition 2.5.20]{BPR10}. 

A {\it positive  Bernoulli distribution} is a morphism $\pi$ from the free monoid $A^*$ onto the multiplicative monoid  $[0,1]$,  such that we have $\pi(a)>0$ for every $a\in A$, and such that $\sum_{a\in A} \pi(a)=1$.
The {\it uniform} distribution corresponds to $\pi(a)=1/|A|$, for every letter $a$. For any subset $X$ of $A^*$, we set $\pi(X)=\sum_{x\in X} \pi(x)$. 
Clearly, the last sum may be finite or not, however  if $X$ is a thin subset we have $\pi(X)<\infty$ \cite[Proposition 2.5.12]{BPR10}; 
moreover  for every code $X\subseteq A^*$, we have $\pi(X)\le 1$.
From this point of view, the following result was  established by  Sh\"utzenberger (e.g. \cite[Theorem 2.5.16]{BPR10}):
\begin{theorem}
\label{classic}
Given a thin code $X\subseteq A^*$, the four following conditions are equivalent:

{\rm (i)} $X$ is complete.

{\rm (ii)}  $X$ is a maximal code.

{\rm (iii)} A positive Bernoulli distribution $\pi$ exists such that $\pi(X)=1$.

{\rm (iv)} For any positive Bernoulli distribution $\pi$, we have $\pi(X)=1$.
\end{theorem}
\subsection{(Anti-)automorphisms}
\label{anti-}
In the whole paper, we fix an alphabet $A$ and a  mapping $\theta$ onto $A^*$
which  is either an {\it automorphism} or an {\it anti-automorphism}: it is an anti-automorphism if it is one-to-one, with $\theta(\varepsilon)=\varepsilon$ 
and  $\theta(xy)=\theta(y)\theta(x)$, for any pair of words $x,y$. For short in any case we write that $\theta$ is an {\it (anti-)automorphism}.

We say that the (anti-)automorphism $\theta$ is  of  {\it finite   order}  if  some positive  integer $k$ exist such that $\theta^k=id_{A^*}$,
the smallest one being the so-called {\it order} of $\theta$ (trivially $id_{A^*}$ is of order $1$). 
It is well known that such a condition is satisfied whenever
$A$ is a finite set; in particular, 
over a two letter alphabet,
any non-trivial (anti-)automorphism is of order $2$ that is, it is {\it involutive}.

In the whole paper, we are interested in the family of sets $X\subseteq A^*$ that are invariant under 
$\theta$ ({\it $\theta$-invariant} for short) that is, which satisfy $\theta(X)\subseteq X$; the mapping $\theta$ being one-to-one,  this is equivalent to $\theta(X)=X$. 
\begin{expl}
\label{ex1}
Let $A=\{a,b,c,d\}$. Consider the (unique) anti-automorphism $\theta$ that is defined by $\theta(a)=a,\theta(b)=b,\theta(c)=d,\theta(d)=c$.
It is straightforward to verify that the mapping $\theta$ is involutive, moreover the sets $\{cd\}$ and $\{abcd,cdba\}$ are $\theta$-invariant.
\end{expl}
\begin{rmk}
\label{Remark1}
{\rm In the spirit of the families of codes that were introduced in \cite{KM06}, given an (anti-)automorphism $\theta$,
define a $\theta$-{\it code} as a set $X$ such that $\bigcup_{i\in{\mathbb Z}}\theta^i(X)$ is a code.
Clearly, with this definition any $\theta$-code is a code; the converse is false, as attested below by  Example \ref{theta-code}.

Actually, any $\theta$-code that is a maximal code, is necessarily $\theta$-invariant.
Indeed, assuming  $X$ not $\theta$-invariant, we have $X\subsetneq X\cup \theta(X)$, thus $X$ is strictly included in the code $\bigcup_{i\in{\mathbb Z}}\theta^i(X)$. 

A similar argument proves that if $X$ is maximal as a $\theta$-code, then it is $\theta$-invariant (indeed, $\bigcup_{i\in{\mathbb Z}}\theta^i(X)$ itself is a $\theta$-code). 

Taking account of the fundamental importance of the concept of maximality in the theory of codes, such properties reinforces the relevance of the notion of $\theta$-invariant code.}
\end{rmk}
\begin{expl}
\label{theta-code}
Let $A=\{a,b\}$ and $\theta$ be the so-called {\it mirror antimorphism}: $\theta(a)=a$, $\theta(b)=b$. Take for $X$ the finite (prefix) code $\{a,ba\}$.
We have $X\cup\theta(X)=\{a,ab,ba\}$, which is not a  code ($ab\cdot a=a\cdot ba$).
\end{expl}
\section{A defect effect for invariant sets}
\label{DEFECT}
We start with some considerations about  $\theta$-invariant submonoids of $A^*$.
Clearly the intersection of a non-empty family of $\theta$-invariant free submonoids of $A^*$ is itself a $\theta$-invariant free  submonoid.
Given a submonoid $M$ of $A^*$, recall that its {\it minimal generating set} is $(M\setminus\{\varepsilon\})\setminus (M\setminus\{\varepsilon\})^2$.
The following property holds:
\begin{proposit}
\label{base}
Given an alphabet $A$ and given an   (anti-)automorphism $\theta$  of $A^*$, let $M$ be a submonoid of $A^*$ and let $S\subseteq A^*$ such that $M=S^*$. 
Then the two following properties hold:

{\rm (i)} If $S$ is $\theta$-invariant then the same holds for $M$.

{\rm (ii)} If $S$ is the minimal generating set of $M$ and 
if $M$ is $\theta$-invariant then $S$ is $\theta$-invariant.
\end{proposit}
\newproof{potA-1}{\bf Proof}
\begin{potA-1}
(i) Assume that the set $S$ is $\theta$-invariant, and let $w\in M$.
Since $M=S^*$,  a finite sequence of words in $S$, namely $(s_i)_{1\le i\le n}$, exists such that $w=s_ 1\cdots s_n$.
Since $\theta$ is an (anti-)automorphism, in every case $\theta(w)$ is some concatenation of  the words $\theta(s_i)$ ($1\le i\le n$), therefore we have  $\theta(w)\in S^*=M$. Consequently $M$ is $\theta$-invariant.

(ii) Assume that $M$ is $\theta$-invariant and let $s\in S$.
It follows from $S\subseteq M$ that we have $\theta(s)\in\theta(M)=M$ therefore, a  sequence of words in $S$, namely $(s_i)_{1\le i\le n}$, exists such that $\theta(s)=s_1\cdots s_n$.
Since $\theta$ is an (anti-)automorphism,  $s$ is in fact some concatenation of the words $\theta^{-1}(s_1),\cdots, \theta^{-1}(s_n)\in M$.
Moreover, for each integer $i\in [1,n]$, we have $\theta^{-1}(s_i)=s_i^1\cdots s_i^{n_i}$, with $s_i^j\in S$ ($1\le j\le n_i$).
It follows from the definition of $S$ that we have $n=1$ and $s=s_1^1=\theta^{-1}(s_1)$, thus $\theta(s)=s_1\in S$.
 As a consequence, $S$ itself is $\theta$-invariant.\cqfd
\end{potA-1}

Informally, the famous defect theorem says that if some words in a set $X$ satisfy a non-trivial equation, then these words can  be written 
upon an alphabet of smaller size. 
In this section, we will examine whether a corresponding result may be stated in the framework of $\theta$-invariant sets.
\begin{theorem}
\label{defect}
Given an alphabet $A$ and given   an (anti-)automorphism $\theta$ of $A^*$, let $X\subseteq A^*$ be a $\theta$-invariant set. Let $Y$ be the minimal  generating set of the smallest $\theta$-invariant free submonoid of $A^*$ 
that contains $X$. If  $X$ is not a code, then we have $|Y|\le |X|-1$.
\end{theorem}
With the notation of Theorem \ref{defect}, since $Y$ is  a code, each word $x\in X$ has a unique factorization upon the words of $Y$, 
namely $x=y_1\cdots y_n$, with $y_i\in Y$ ($1\le i\le n$). 
In a classical way, we say that $y_1$ ($y_n$)  is the {\it initial} ({\it terminal}) factor of $x$ (with respect to such a factorization). 
From this point of view, before to prove Theorem \ref{defect}, we need to establish the following statement:
 \begin{lem}
\label{initial}
With the preceding notation, each word in $Y$ is the initial (terminal) factor of some word in $X$.
 \end{lem}
\newproof{potA0}{\bf Proof}
\begin{potA0}
By contradiction, assume that a word $y\in Y$  that is never initial of any word in $X$ exists. 
Set $Z_0=(Y\setminus \{y\})\{y\}^*$ and  $Z_i=\theta^i(Z_0)$, for each integer $i\in {\mathbb Z}$. 
In a classical way (see e.g. \cite[p.  7]{Lo1983}), since $Y$ is a code, $Z_0$ itself is a code.\\
For each integer $i\in  {\mathbb Z}$, since $\theta^i$ is itself an (anti-)automorphism,  $Z_i$ is a code that is, $Z_i^*$ is a free submonoid of $A^*$. 
Consequently, the intersection, say $M$, of the family $(Z_i^*)_{i\in {\mathbb Z}}$ is itself a free submonoid of $A^*$. \\
Let $w\in M$. For each integer $i\in {\mathbb Z}$, we have $w\in Z_i^*$, thus $\theta(w)\in \theta(Z_i^*)\subseteq (\theta(Z_i))^*= (\theta^{i+1}(Z_0))^*=Z_{i+1}^*$.
Consequently we have $\theta(w)\in\bigcap_{i\in {\mathbb Z}}Z_{i+1}^*=\bigcap_{i\in {\mathbb Z}}Z_{i}^*=M$,
whence we have $\theta(M)\subseteq M$  therefore, since $\theta$ is onto, we obtain $\theta(M)=M$.\\
Let $x$ be  an arbitrary word in  $X$. Since $X\subseteq Y^*$,  and according to  the definition of $y$, 
we have $x=(z_1y^{k_1})(z_2y^{k_2})\cdots (z_ny^{k_n})$, with $n\ge 1$, $z_1,\cdots z_n\in Y\setminus\{ y\}$ and $k_1,\cdots k_n\ge 0$. 
Consequently $x$ belongs to $Z_0^*$, therefore we have $X\subseteq Z_0^*$. 
Since  $X$ is $\theta$-invariant, this implies $X=\theta^i(X)\subseteq \theta^i(Z_0^*)\subseteq Z_i^*$, for each $i\in {\mathbb Z}$, thus $X\subseteq M$.\\
But the word $y$ belongs to $Y^*$ and does not belong to $Z_0^*$  thus, it doesn't belong to $M$.  
This implies $X\subseteq M\subsetneq Y^*$: a contradiction with the minimality of $Y^*$. Clearly, similar arguments may be applied to words $y\in Y$ that are never terminal of any word in $X$: this completes the proof.\cqfd
\end{potA0}

\newproof{potAB}{\bf Proof of Theorem \ref{defect}}
\begin{potAB}
 Let $\alpha$ be the mapping from  $X$ onto $Y$ which, with every word $x\in X$, 
 associates  the initial factor of 
$x$ in its (unique) factorization over $Y^*$. According to Lemma \ref{initial}, $\alpha$ is onto.
We will prove that it is not one-to-one. Classically,  since $X$ is not a code, a non-trivial equation may be written among its words, say:
$x_1\cdots x_n=x'_1\cdots x'_p,~~ {\rm with}~~x_i,x'_j\in X~~x_1\neq x'_1~~(1\le i\le n, 1\le j \le p).$
Since $Y$ is a code, a unique sequence of words in $Y$, namely $y_1,\cdots, y_m$ ($m\ge 1$) exists such that:
$x_1\cdots x_n=x'_1\cdots x'_p=y_1\cdots y_m. $
 This implies $y_1=\alpha(x_1)=\alpha(x'_1)$ and completes the proof. \cqfd
\end{potAB}
In what follows we discuss some  interpretation of Theorem \ref{defect} with regard to equations in words.
For this purpose, we assume that $A$ is finite, $\theta$ being of order $k$, and we 
consider  a finite set of words, say $Z$.  Let $X$ be the union of the sets $\theta^i(Z)$, for $i\in [0,k-1]$, and assume that a non-trivial equation holds among the words of $X$, namely
$x_1\cdots x_m=y_1\cdots y_p$.
By construction $X$ is $\theta$-invariant therefore, 
according to Theorem \ref{defect}, a $\theta$-invariant code $Y$ exists such that $X\subseteq Y^*$, with  $|Y|\le |X|-1$. 
This means that each of the words in $X$ can be expressed by making use of at most $|X|-1$ words of type $\theta^i(u)$, 
with $u\in Y$ and $0\le i\le k-1$.
It will be easily verified that the examples from \cite{CCKS11,KM08,MMNS14} corroborate this fact; moreover, below  
we mention an original  one: 
\begin{expl}
\label{ex1}
Let  $\theta$ be an anti-automorphism of order $3$.
Consider two different  words $x,y$, with $|x|>|y|>0$, satisfying the equation:
$ x\theta(y)=\theta^2(y)\theta(x).$
With this condition, a pair of words $u,v$ exists such that $x=uv$,  $\theta(x)=v\theta(y)$, $\theta^2(y)=u$,
thus $y=\theta(u)$. It follows from $x=uv$ that $v\theta(y)=\theta(x)=\theta(v)\theta(u)$, thus $v=\theta(v)$ and $\theta(y)=\theta(u)$. 
This implies $y=u=\theta(u)=\theta^2(u)$ and $v=\theta(v)=\theta^2(v)$.
Moreover, we have  $\theta(x)=vu$, $\theta^2(x)=uv$: we obtain $x=\theta^2(x)$ thus,  $x=\theta(x)=\theta^2(x)$; hence we have $uv=vu$.
Consequently, a non-empty word $t$ and  integers $i,j$ exist such that $u=t^i$, $v=t^j$.
With the preceding notation, we have $Z=\{x,y\}$, $X=Z\cup\theta(Z)\cup\theta^2(Z)=\{x,y\}$, $Y=\{t\}$.
We verify that  $|Y|\le |X|-1$.
\end{expl}

\section{Finite complete $\theta$-invariant codes}
\label{finite-comp}
In this section we are interested in finite  complete $\theta$-invariant codes over an alphabet $A$.
Given an arbitrary letter $a\in A$, since for every non-negative integer $n$, we  have $a^n\in F(X^*)$, necessarily a (unique) positive integer $p$ exists such that $a^p\in X$;
therefore, $A$ is necessarily finite. 
Several examples of finite  complete $\theta$-invariant codes will be presented.
We start with prefix codes, which certainly constitute the best-known class of them.
\subsection{Finite complete prefix $\theta$-invariant codes}
Actually finite complete prefix codes play a peculiar part in the framework of codes. A famous result due to Sch\"utzenberger \cite{S66} (cf. also \cite{Br92})
states that any finite complete code with a {\it finite deciphering delay} (e.g. \cite[Ch. 5]{BPR10}) is necessarily  a prefix code.
In particular, over $A^*$ only one finite complete {\it circular} code  (or, equivalently, finite complete {\it uniformly synchronized code}) can exist, namely the alphabet $A$ itself (cf. \cite[Ch. 7, Ch. 10]{BPR10},  \cite{L76}).

It is well-known that each prefix set, say $X$, can be represented by a tree, say ${\cal T}(X)$,  of arity $|A|$:
 in this representation, each node (i.e. vertice) is a prefix of some word in $X$ (i.e. an element of $P(X)$),
the root being $\varepsilon$, the empty word. Moreover, given two nodes $u$, $v$ and a letter $a\in A$,  an 
edge with label $a$  exists from $u$ to $v$ in  ${\cal T}(X)$ if, and only if, we have $v=ua$: we denote such a labeled edge by $(u,a,v)$ and we say that $v$ is a successor of $u$. 
In that representation, complete prefix codes correspond to {\it complete trees}, in the sense where each interior node has exactly  $|A|$ successors. 

We start with the case where  $\theta$ is an automorphism of $A^*$. Given a prefix set $X\subseteq A^*$, we say that the corresponding tree  ${\cal T}(X)$ is {\it  invariant} under $\theta$ whenever
$(u,a,v)$ is an edge of ${\cal T}(X)$ if, and only  if,  $(\theta(u),\theta(a),\theta(v))$ is an edge of ${\cal T}(X)$.
With this notion, a characterization of $\theta$-invariant  prefix codes may be stated:
\begin{fait}
\label{fact-000}
Let $A$ be a finite alphabet, let $\theta$ be an automorphism of $A^*$ and
let $X$ be a  prefix code. Then $X$ is $\theta$-invariant if, and only if, the tree  ${\cal T}(X)$ itself is 
invariant under $\theta$. 
\end{fait}
\newproof{pot-000}{\bf Proof}
\begin{pot-000}
Assume that $X$ is $\theta$-invariant, and let  $(u,a,ua)$ an arbitrary edge in ${\cal T}(X)$. By construction a  word $s\in S(X)$ exists such that  $uas\in X$.
Since $X$ is a $\theta$-invariant set, this implies  $\theta(u)\theta(as)=\theta(ua)\theta(s)\in X$, thus $\theta(u)$ and $\theta(ua)\in P(X)$.
Consequently, $(\theta(u),\theta(a),\theta(u)\theta(a))$ is an edge of ${\cal T}(X)$, therefore  ${\cal T}(X)$  is invariant under $\theta$.

Conversely, assume that   ${\cal T}(X)$ is invariant under $\theta$. Let $w=w_1\cdots w_n\in X$, with $w_i\in A$ ($1\le i\le n$). 
By construction,  the following sequence  of edges exists in ${\cal T}(X)$ (for $i=0$, we set $w_1\cdots w_{i}=\varepsilon$):
$$(w_1\cdots w_{i},w_{i+1},w_1\cdots w_{i+1})~~~~(0\le i\le n-1),$$
moreover the node $w=w_1\cdots w_{n}$ has no successor.
Since ${\cal T}(X)$ is invariant under $\theta$, a corresponding sequence of edges exists in ${\cal T}(X)$, 
namely:
$$ (\theta(w_1)\cdots \theta(w_{i}),\theta(w_{i+1}), \theta(w_1)\cdots \theta(w_{i+1}))~~~~(0\le i\le n-1).$$
Since the node $w_1\cdots w_{n}$ has no successor, the same holds for the corresponding node $\theta(w)=\theta(w_1\cdots w_{n})$: this implies $\theta(w)\in X$. 
\cqfd
\end{pot-000}
\begin{expl}
\label{E00}
Let $A=\{a,b\}$, and $\theta$ be the automorphism defined by $\theta(a)=b$, $\theta(b)=a$.
Given an arbitrary integer $n\ge 3$,
consider the following set: 
$$X=\bigcup_{1\le i\le n-1} \{a^i b, b^ia\}\cup \{a^n, b^n\}.$$
By construction, $X$ is a prefix code.  Moreover, $X$ is complete: this can be directly verified by examining ${\cal T}(X)$ (an alternative method consists in applying
Theorem \ref{classic} (iii), with $\pi$ the uniform Bernoulli distribution).
It is also straightforward to verify that  $X$ is $\theta$-invariant. 

Note that $X$ is not bifix: indeed, for each integer $i\in [2,n-1]$, the word $ab\in X$ is a suffix of  $a^ib\in X$.
Figure 1 illustrates the corresponding  tree ${\cal T}(X)$ for $n=4$.
\end{expl}
\begin{figure}
\begin{center}
\includegraphics[width=9.75cm,height=12cm]{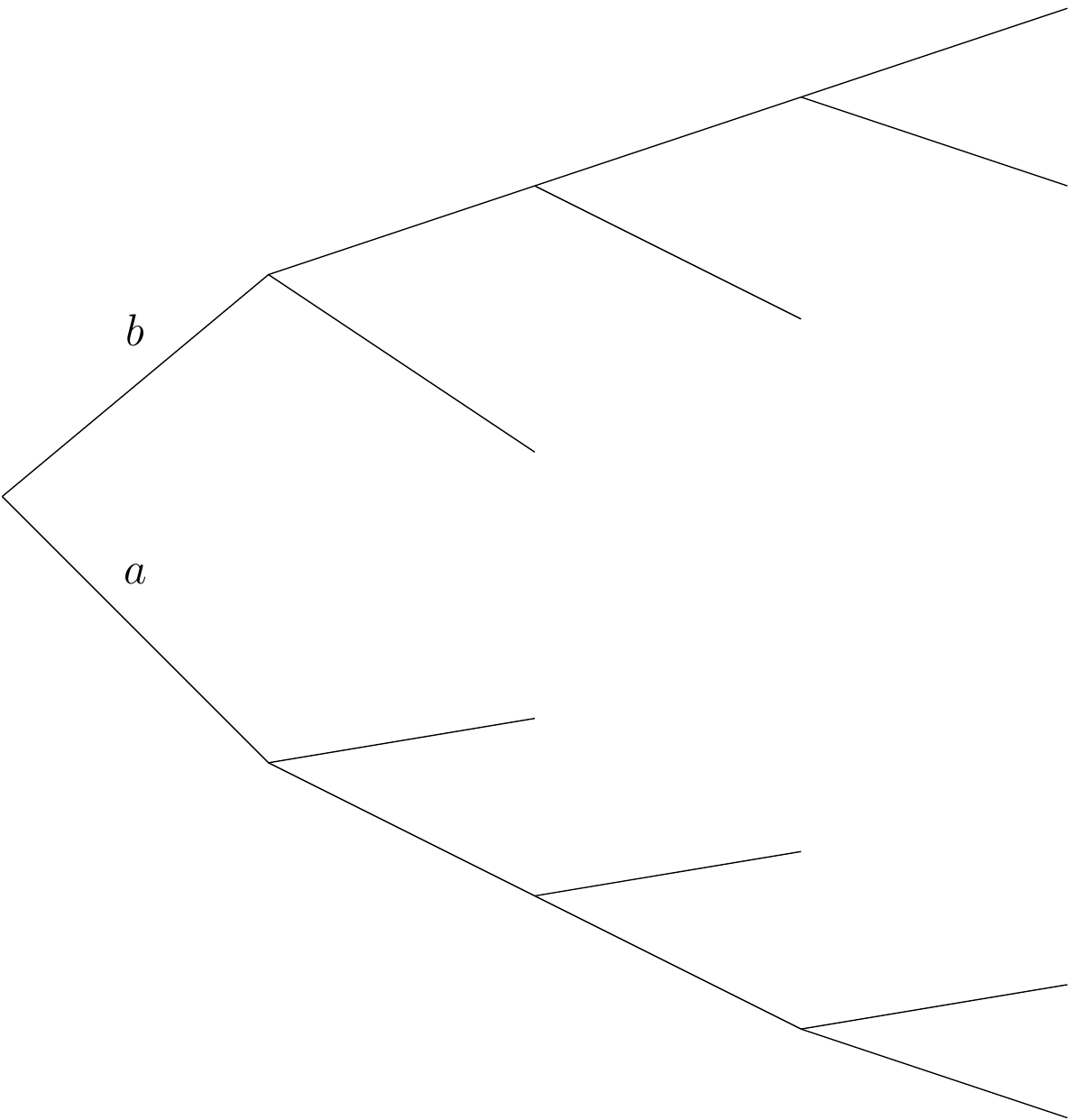}
\end{center}
\caption[]
{Example \ref{E00} with $n=4$. In the tree ${\cal T}(X)$, each bottom-up (top-down) branch represents an edge with label $b$ ($a$).}
\end{figure}
In the case where $\theta$ is an anti-automorphism, the following property is noticeable:
\begin{fait}
\label{fact0}
Let $\theta$ be an anti-automorphism onto $A^*$ and let $X\subseteq A^*$ be a finite $\theta$-invariant code.
If $X$ is prefix, then it is necessarily bifix.
\end{fait}
\newproof{pot00}{\bf Proof}
\begin{pot00}
By contradiction, assume $X$ not bifix, thus not suffix: words $p\in A^*$, $s\in A^+$ exist such that $s,ps\in X$.
Since $X$ is $\theta$-invariant, we have $\theta(s),\theta(ps)\in X$, thus $\theta(s),\theta(s)\theta(p)\in X$: this contradicts the fact that $X$ is a prefix code.\cqfd
\end{pot00}
The result of Claim \ref{fact0} directly leads to examine the behavior of finite complete bifix codes with regard to (anti-)automorphisms.
\subsection{Finite complete  bifix  $\theta$-invariant codes}
At first, it is worth mentioning a well-known class of finite bifix codes: 
\begin{expl}
A set $X$ is {\it  uniform} if a positive integer $n$ exists such that $X\subseteq A^n$. 
Trivially, such a set is a bifix code moreover, it is complete if, and only if, we have $X=A^n$.
It is straightforward to verify that $X$ is invariant under every (anti-)automorphism of $A^*$: 
indeed, the restriction of such a mapping on words of length $n$ induces a permutation of $A^n$.
\end{expl}
It is a natural question to ask whether non-uniform finite complete bifix $\theta$-invariant  codes exist.
By exhibiting infinite classes of convenient codes, the three  following  examples allow to bring a positive answer. 
Actually, the two first families of codes have been constructed by applying a famous internal transformation to some uniform code \cite{C72} (cf. also \cite[§     6.2]{BPR10})
\begin{expl}
\label{E2}
Let $A=\{a,b\}$ and  $\theta$ be  the anti-automorphism of $A^*$ that is defined by $\theta(a)=b$, $\theta(b)=a$.\\
Let  $n=2k+1$, with $k\ge 1$. Consider the following set:
$$X=(A^n\setminus (Aa^kb^k\cup a^kb^kA))\cup\{a^kb^k\}\cup Aa^kb^kA.$$
The set $Aa^kb^kA$ is a (uniform) bifix code.
Since the condition  $a^kb^k\in P(X)$ ($a^kb^k\in S(X)$) necessarily implies $a^kb^k \in P(a^kb^kA)$ ($a^kb^k\in S(Aa^kb^k)$),
$X$ is a finite (non-uniform) bifix code.
The code $X$ is complete: indeed, we have $Aa^kb^k\cap a^kb^kA=\emptyset$ therefore, 
given an arbitrary positive Bernoulli distribution $\pi$ over $A^*$, we have:   
$$\pi(X)=\pi(A^n\setminus (Aa^kb^k\cup a^kb^kA))+\pi(a^kb^k)+\pi(Aa^kb^kA)=1-2\pi({a^kb^k})+2\pi(a^kb^k)=1.$$ 
Furthermore, since we have $\theta(A)=A$ and $\theta(a^kb^k)=a^kb^k$, $X$ is $\theta$-invariant.\\
For $n=3$, the preceding construction leads to the following  finite complete bifix $\theta$-invariant  code \cite[(1)]{C72}:
 $$X=\{a^3, ba^2, b^2a, b^3, ab, a^2ba,a^2b^2, baba, bab^2\}.$$
\end{expl}
\begin{expl}
\label{E21}
Let $A=\{a,b\}$ and $\theta$ be the so-called mirror-image, which is in fact  the anti-automorphism defined by $\theta(a)=a$, $\theta(b)=b$.\\ 
Take $n=3k+1$, with $k\ge 1$. We have $a^kb^ka^k\not\in P(Aa^kb^ka^k)\cup S(a^kb^ka^kA)$; therefore
an examination similar to the one we applied at Example \ref{E2} leads to verify that the following set is a  finite complete bifix $\theta$-invariant code:
$$X=(A^n\setminus (Aa^{k}b^{k}a^{k}\cup a^{k}b^{k}a^{k}A))\cup\{a^{k}b^{k}a^{k}\}\cup Aa^{k}b^{k}a^{k}A.$$
For $n=4$ (i.e. $k=1$) the corresponding binary tree ${\cal T}(X)$ is represented in Figure 2.
\end{expl}
We observe that, in view of constructing arbitrarily large non-uniform finite bifix $\theta$-invariant  codes over arbitrarily large finite alphabets, the two last constructions can be generalized,
as illustrated by the following example: 
\begin{expl}
\label{E22}
1) Let $A=\{a,b,c\}$ and $\theta$ be the  anti-automorphism defined by $\theta(a)=b$, $\theta(b)=c$, $\theta(c)=a$.

Take $n=2k+1$, with $k\ge 1$, and $$W=\bigcup_{i\in{\mathbb Z}}\{\theta^i(a^kb^k)\}=\{a^kb^k,c^kb^k,c^ka^k, b^ka^k,b^kc^k,a^kc^k\}.$$ 
By construction we have $W\cap (P(AW)\cup S(WA))=\emptyset$ therefore,  
the following set is a $\theta$-invariant finite bifix code:
\begin{equation}
\label{EQ}
X=(A^n\setminus (AW\cup WA))\cup W\cup AWA.
\end{equation}
Moreover, $X$ is complete: indeed, by construction we have $AW\cap WA=\emptyset$ therefore, for any positive Bernoulli distribution over $A^*$, we have $\pi(X)=1-2\pi(W)+2\pi(W)=1$.\\

2) Similarly,  over $A=\{a,b,c\}$ take for $\theta$ the anti-automorphism onto $A^*$ defined  by $\theta(a)=b$, $\theta(b)=c$, $\theta(c)=a$.
Set  $n=3k+1$ and $$W=\bigcup_{i\in{\mathbb Z}}\{\theta^i(a^kb^ka^k)\}=\{a^kb^ka^k, b^kc^kb^k, c^ka^kc^k\}.$$
Applying the construction from (\ref{EQ}) also leads 
to  obtain a finite complete bifix $\theta$-invariant  code.
\end{expl}
\begin{figure}
\begin{center}
\label{Figure9}
\includegraphics[width=9cm,height=9.55cm]{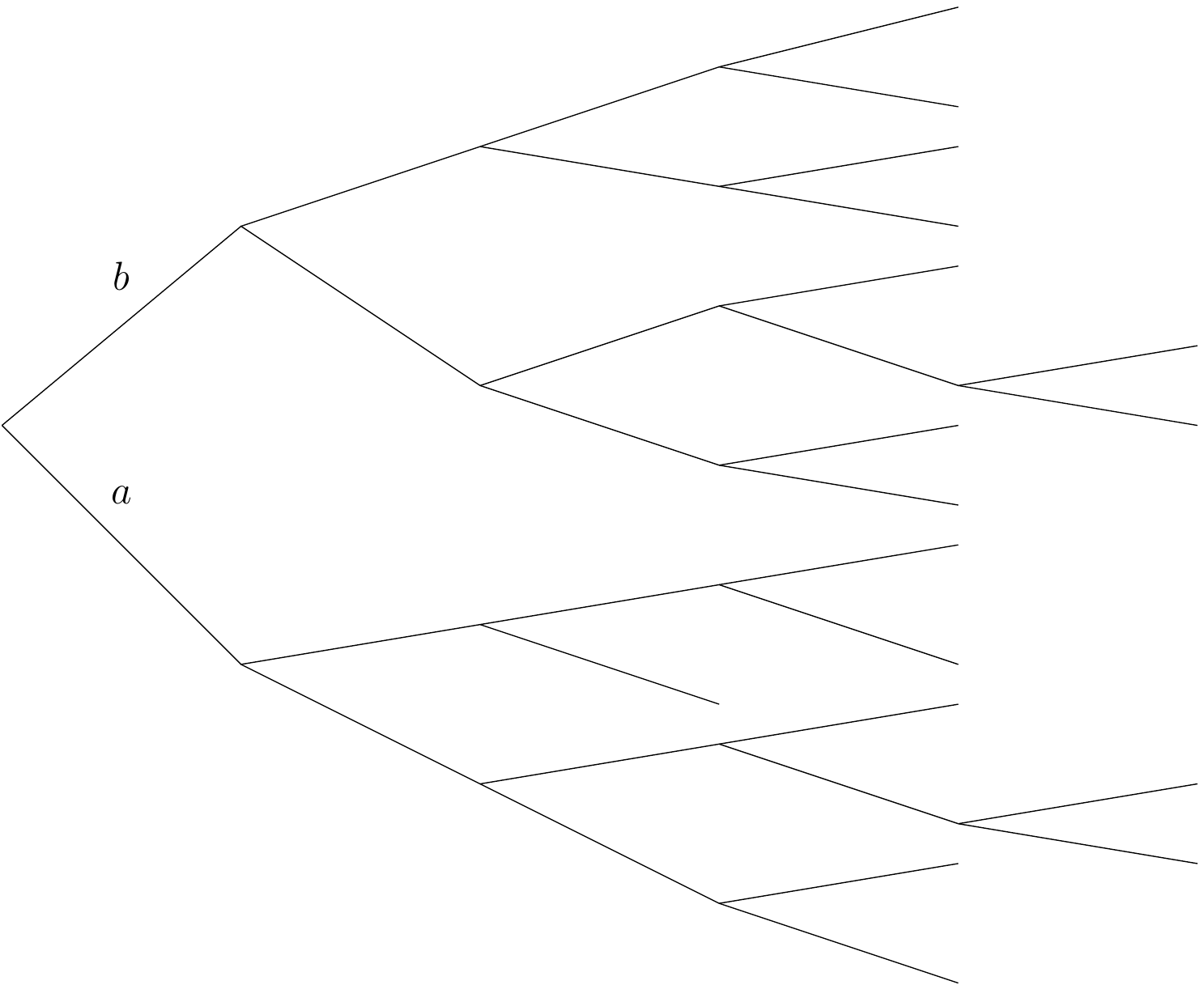}
\end{center}
\caption[]
{Example \ref{E21}: the case where $n=4$, thus $k=1$.}
\end{figure}
\subsection{Non-prefix non-suffix finite complete $\theta$-invariant codes}
\label{general}
In the most general case, given an (anti-)automorphism $\theta$ of $A^*$,
we are looking to finite codes $X\subseteq A^*$ which are both complete and $\theta$-invariant.
With regard to the last condition, the following statement brings some characterization: 
\begin{fait}
\label{fact1}
Let $\theta$ be an (anti-)automorphism onto $A^*$ and let $X$ be a finite subset of $A^*$.
Then   $X$ is $\theta$-invariant if, and only if, it is the disjoint union of a finite family of uniform $\theta$-invariant codes.
\end{fait}
\newproof{prf1}{\bf Proof}
\begin{prf1}
 Let $\ell_1<\cdots<\ell_n$ be the unique increasing finite sequence of the lengths of the words in $X$. For each  $i\in [1,n]$, set $X_{i}=X\cap A^{\ell_i}$.
By construction, each set $X_{i}$ is a uniform code, moreover we have:
$$X=\bigcup_{1\le i\le n} X_i.$$ Clearly, the set $X$ is $\theta$-invariant if, and only if,
for each integer $i\in [1,n]$, $\theta$ induces a permutation of $X_i$ itself.\cqfd 
\end{prf1}
When $X$ is a required to be a code, Claim \ref{fact1} only   leads to some necessary condition. 
For instance, the set $\{a,ab,b\}=\{a,b\}\cup\{ab\}$, which satisfies the condition of the claim, is $\theta$-invariant, but clearly  it is not a code. 
Actually, despite that in any case $\theta$-invariance is preserved with respect to the union of sets, the main obstacle is that, given two (disjoint) codes, there is no characterization that can guarantee that their union remains a code. 
 
Of course, one can wonder about the impact of $\theta$-invariance itself on the structure of a finite complete code.
Indeed, in view of the above, such an influence is very strong with regard to two special families of codes: the uniform ones and, with respect to automorphisms, the family of prefix non-suffix codes. 
However, the part of $\theta$-invariance  appeared in fact of lesser importance in the construction of our  families of bifix codes, where it essentially involved the structure of a few  convenient words (e.g. the elements of $W$). 

Things become even more complex when attempting to construct finite  complete $\theta$-invariant codes that are neither prefix nor suffix. 
Indeed, with regard to finite complete codes, although that some famous families  have been exhibited (e.g. \cite{DF89,DFR85}), no general  structure is known.
However, finite complete $\theta$-invariant codes that are neither  prefix  nor suffix  exist as attested by the  following example:
\begin{expl} 
\label{E3}
With the anti-automorphism $\theta$ that was introduced in Example \ref{E2} (which swaps the letters $a$ and $b$), consider the classical finite complete code
$X=\{a^2,ab,a^2b,ab^2, b^2\}$ \cite[Example 2]{RSS89}, which is neither prefix, nor suffix.
It is straightforward to verify that it is $\theta$-invariant (we have $\theta(ab)=ab$). 
\end{expl}
\subsection{\it Toward the construction of regular complete $\theta$-invariant codes}
In \cite{R75}, by making use of factorizations of the so-called cyclotomic
polynomials, the author provided a family of non-finitely completable codes.
It is therefore a natural question to ask whether corresponding objects exist in the framework of $\theta$-invariant codes.

Let $A$ be a finite alphabet, and let $\theta$ be an (anti)-automorphism of $A^*$.
Given a finite code $X$, if $X$ is embeddable into a complete $\theta$-invariant code, say $Y$,   then,
with the terminology of Remark \ref{Remark1}, it has to be a $\theta$-code.
Indeed  the set $\bigcup_{i\in{\mathbb Z}}\theta^i(X)$ is necessarily a $\theta$-invariant code that is included in $Y$. 
Therefore, our problem comes down to wonder whether a given finite $\theta$-invariant code can be embedded into a complete one.

We begin by strictly restraining  the problem to the framework of prefix codes.
Given  a (non-trivial) automorphism $\theta$, according to the preceding Claim \ref{fact-000} any $\theta$-invariant prefix code can be embedded into a  $\theta$-invariant complete one.
Informally, if suffices to complete the corresponding tree with convenient ones of arity  $|A|$ that are invariant under $\theta$.

In the case of anti-automorphisms, according to Claim \ref{fact0}, for being embeddable into a complete  one, a finite prefix $\theta$-invariant  code has to be bifix.
However, the converse is false;  indeed there are  finite  bifix  $\theta$-invariant  codes that cannot be included into any complete one, as attested by the following example: 
\begin{expl}
\label{E33}
1) Let $A=\{a,b\}$, and be $\theta$ be the  mirror anti-automorphism of Example \ref{E21}.\\
At first, we observe that the finite $\theta$-invariant bifix code $X=\{aa,b\}$ cannot embed into any finite complete  bifix  (not necessarily $\theta$-invariant) code. 
Indeed, assume that such a complete code, say $Z$, exists: necessarily $Z$ is prefix and complete, hence for any positive integer $p$, we have $ab^p\in P(Z^*)$. Therefore a positive integer $n$ exists such that $ab^n$ belongs to $Z$;
since $b$ belongs to $Z$ this contradicts the fact that $Z$ is bifix.\\ 
As a consequence $X$ cannot be included in any finite complete  prefix  $\theta$-invariant code. Indeed, according to Claim \ref{fact0}, such a code should be bifix.

2) Note that the infinite (regular) set $Z= \{b\}\cup\{ab^na: n\in {\mathbb N}\}$ is a $\theta$-invariant bifix code which contains $X$.
Moreover, taking for $\pi$ the  uniform Bernoulli distribution, it is straightforward to verify that  we have $\pi(Z)=1/2+1/4\sum_{n\in  {\mathbb N}}(1/2^n)=1$, thus $Z$ is complete. 
\end{expl}
We do not know whether there are  finite $\theta$-invariant  complete codes that contain the code $X$ of Example \ref{E33}.
Actually, as far as we know, the question of embedding a finite $\theta$-invariant code into a complete one remains open.

From another angle, the study in \cite{R75}  led its author to conclude that the study of all finite codes requires also investigations on the infinite ones. 
From that, the question of embedding a finite code into a regular one was open.
 A positive answer was given in \cite{ER85}, where a now famous method for embedding a regular code into a complete one was published.

From this last point of view, in the next section, we will interest in  the problem of embedding a  regular $\theta$-invariant code into a regular complete one.

\section{Embedding a regular $\theta$-invariant code into a complete one}
\label{embedding}
\subsection{Some notation}
In this section we consider  an (anti-)automorphism $\theta$ of $A^*$, and a  non-complete  $\theta$-invariant code $X\subseteq A^*$.
We ask for a  complete regular  $\theta$-invariant code $Y$ such that $X\subseteq Y$.
We will bring  a positive answer: let's begin by describing our construction.

Let $X$ be a non-complete $\theta$-invariant code, and let $y\not\in F(X^*)$. Necessarily, we have $|A|\ge 2$ (otherwise, $X$ should be  complete). 
Without loss of generality, 
we may assume that the initial and the terminal letters of $y$ are different (otherwise, substitute to $y$ the word $ay{\overline a}$, 
with $a,{\overline a}\in A$ and $a\neq {\overline a}$): in particular, we have $|y|\ge 2$. 
Set:
\begin{eqnarray}
\label{Eq2}
 z={\overline a^{|y|}}ya^{|y|}~~~({\rm with}~~y\in a A^*{\overline a}).
\end{eqnarray}
Since $\theta$ is an (anti-)automorphism, for each integer $i\in {\mathbb Z}$, two different letters  $b,{\overline b}$ 
 exist such that the following property holds:
\begin{eqnarray}
\label{Eq3}
\theta^i(z)={\overline b^{|y|}}\theta^i(y)b^{|y|}~~~({\rm with}~~\theta^i(y)\in b A^*{\overline b}).
\end{eqnarray}
Finally, we introduce the three following sets:
\begin{eqnarray}
\label{Z}
Z=\bigcup_{i\in{\mathbb Z}}\{\theta^i(z)\},
\end{eqnarray}
\begin{eqnarray}
\label{W}
W=ZA^*\cap A^*Z,
\end{eqnarray}
\begin{eqnarray}
\label{T}
T=W\setminus (W\cup X)(W\cup X)^+.
\end{eqnarray}
By construction,  the following inclusion holds:
\begin{eqnarray}
\label{decomposable}
W\subseteq (X\cup T)^+.
\end{eqnarray}
\subsection{Basic properties of $Z$}
\label{BASIC}
By construction, each element of the  preceding set $Z$ has length  $3|y|$.
Given two (not necessarily different) integers $i,j\in {\mathbb Z}$, we will accurately study how the two words $\theta^i(z), \theta^j(z)$ may overlap. 
 \begin{lem}
\label{overlapZ}
With the notation in (\ref{Eq3}), let $u,v\in A^+$, $i,j\in {\mathbb Z}$ such that  $|u| \le |z|-1$ and $\theta^i(z)v=u\theta^j(z)$. 
Then we have $|u|=|v|\ge 2|y|$, moreover a letter $b$ and a unique positive integer $k$ (depending of $|u|$) exist such that we have $\theta^i(z)=ub^k$, $\theta^j(z)=b^kv$, with $k\le |y|$. 
 \end{lem}
\newproof{pot01}{\bf Proof}
\begin{pot01}
According to (\ref{Eq3}), we set 
$\theta^i(z)={\overline b^{|y|}}bx'{\overline b}b^{|y|}$
and 
$\theta^j(z)={\overline c^{|y|}}cx''{\overline c}c^{|y|}$,
with $b, {\overline b}, c, {\overline c}\in A$, $b \neq {\overline b}, c \neq {\overline c}$ and $|x'|=|x"|=|y|-2$. 
Since $\theta$ is an (anti-)automorphism, we have $|\theta^i(z)|=|\theta^j(z)|$, thus $|u|=|v|$; 
since we have $1\le|u|\le 3|y|-1$, exactly one of the following cases occurs:\\
{\it Case 1:} $1\le |u|\le |y|-1$. With this condition, we have 
($\theta^i(z))_{|u|+1}={\overline b}={\overline c}=(u\theta^j(z))_{|u|+1}$ and 
$(\theta^i(z))_{|y|+1}=b={\overline c}=(u\theta^j(z))_{|y|+1}$,
which contradicts $b\neq {\overline b}$.\\
{\it Case 2:} $|u|=|y|$. This condition implies 
$(\theta^i(z))_{|u|+1}=b={\overline c}=(u\theta^j(z))_{|u|+1}$ and 
$(\theta^i(z))_{2|y|}={\overline b}={\overline c}=(u\theta^j(z))_{2|y|}$, 
which contradicts $b\neq {\overline b}$.\\
{\it Case 3:} $|y|+1\le |u|\le 2|y|-1$. We obtain 
$(\theta^i(z))_{2|y|}={\overline b}={\overline c}=(u\theta^j(z))_{2|y|}$ and
$(\theta^i(z))_{2|y|+1}=b={\overline c}=(u\theta^j(z))_{2|y|+1}$
which contradicts $b\neq {\overline b}$.\\
{\it Case 4:} $2|y|\le|u|\leq |z|-1=3|y|-1$. With this condition, necessarily we have  
$(\theta^i(z))_{|u|+1}=b={\overline c}=(u\theta^j(z))_{|u|+1}$,
 therefore an integer $k\in [1,|y|]$ exists such that $\theta^i(z)=ub^k$ and $\theta^j(z)=b^kv$.\cqfd
\end{pot01}
%
%
\begin{lem}
\label{facteurInterne}
With the preceding notation, we have $A^+ZA^+\cap ZX^*Z=\emptyset$.
\end{lem}
\newproof{pot02}{\bf Proof}
\begin{pot02}
By contradiction, assume that $z_1,z_2,z_3\in Z$ , $x\in X^*$ and $u,v\in A^+$ exist such that $uz_1v=z_2xz_3$.
By comparing the lengths of $u,v$ with $|z|$, exactly one of the three following cases occurs:\\
{\it Case 1:} $|z| \le |u|$ and $|z| \le|v|$. With this condition,  we have $z_2\in P(u)$ and $z_3\in S(v)$, therefore 
the word $z_1$ is a factor of $x$: this contradicts $Z\cap F(X^*)=\emptyset$.\\
{\it Case 2:} $|u|<|z|\le |v|$. We have in fact $u\in P(z_2)$ and $z_3\in S(v)$.
We are in the condition of Lemma  \ref{overlapZ}: the words $z_2$, $z_1$  overlap.  
Consequently, $u,z'_1\in A^+$ and $b\in A$ exist such that $z_2=ub^k$ and $z_1=b^kz'_1$, with $1\le k\le |y|$ and $|z'_1|=|u|$. 
But, by construction,  we have $|uz_1|=|z_2xz_3|-|v|$. 
Since we assume $|v|\ge |z|$,  this implies
 $|uz_1|\le|z_2xz_3|-|z|=|z_2x|$, hence we  obtain $uz_1=ub^{k}z'_1\in P(z_2x)$. 
It follows from $z_2=ub^{k}$ that $z'_1\in P(x)$. 
Since we have $z_1\in Z$ and according to (\ref{Eq3}), $i\in{\mathbb Z}$ and ${\overline b}\in A$ exist such that we have $z_1=b^kz'_1=b^{|y|}\theta^i(y){\overline b}^{|y|}$.
Since by Lemma  \ref{overlapZ} we have $|z_1'|=|u| \geq 2|y|$, we obtain
$\theta^i(y) \in F(z'_1)$, thus $\theta^i(y) \in F(x)$, which contradicts $y \notin F(X^*)$.\\
{\it Case 3:} $|v|<|z|\le |u|$. Some similar arguments on the reversed words lead to a conclusion similar to that of  Case 2.\\
{\it Case 4:} $|z|>|u|$ and $|z|>|v|$. With this condition, both the pairs of words $z_2,z_1$ and $z_1,z_3$ overlap.  
Once more we are in the condition of Lemma \ref{overlapZ}: letters $c,d$, 
words $u, v, s,t$, and integers $h,k$ exist such that the two following properties hold:
\begin{eqnarray}
z_2=uc^h,~~  z_1=c^{h}s,~~ |u|=|s| \ge 2|y|, ~~h \le |y|,
\\
z_1=td^{k},~~ z_3=d^kv, ~~ |v|=|t| \ge 2|y|, ~~k \le |y|.
\end{eqnarray}
It follows from $uz_1v=z_2xz_3$ that $uz_1v=(uc^h)x(d^kv)$, thus  $z_1=c^{h}xd^{k}$.
But according to (\ref{Eq3}), $i\in{\mathbb Z}$ and ${\overline c}\in A$ exist such that we have $z_1=c^{|y|}\theta^i(y){\overline c}^{|y|}$.
Since we have $h,k\le |y|$, this implies $d={\overline c}$
moreover $\theta^i(y)$ is a factor of $x$. 
Once more, this contradicts $y \notin F(X^*)$.\cqfd
\end{pot02}

%
%
As a direct consequence of Lemma  \ref{facteurInterne}, we obtain the following result:
\begin{coro}
\label{X-pref}
With the preceding notation, $X^*Z$ is a prefix code.
\end{coro}
\newproof{pot03}{\bf Proof}
\begin{pot03}
Let $z_1,z_2\in Z$, $x_1,x_2\in X^*, u \in A^+$, such that $x_1z_1u= x_2z_2$. For any word $z_3\in Z$, 
we have  $(z_3x_1)z_1(u)=z_3x_2z_1$, a contradiction with Lemma \ref{facteurInterne}.\cqfd
\end{pot03}
\subsection{The consequences for the set $X\cup T$}
\begin{lem}
\label{TcupXcode}
The set $X\cup T$ is a $\theta$-invariant code.
\end{lem}
\newproof{pot1}{\bf Proof}
\begin{pot1}
The fact that $X\cup T$ is $\theta$-invariant comes from its construction. 
For proving that it is a code, we consider an arbitrary equation among the words in $X\cup T$. 
Since $X$ is a code, and since $z\notin F(X^*)$, we may assume that at least one occurrence of a word in $T$
 appears in each side of the equation, therefore this equation takes the following form:
\begin{equation}
x_0t_0 x_1t_1\cdots t_{n-1}x_n=x'_0t'_0\cdots t'_{p-1}x'_p,
\end{equation}
with $x_i,x'_j\in X^*$ $(0\le i\le n$, $0\le j\le p$) and $t_i,t'_j\in T$ ($0\le i\le n-1$, $0\le j\le p-1$).
Since by construction we have $T\subseteq W \subseteq ZA^*$, each side of the equation has a prefix in $X^*Z$.
According to  Corollary \ref{X-pref} and since all the words in $Z$ have a common length,
this implies $x_0=x'_0$, therefore our equation is equivalent to:
\begin{equation}
\label{equation1}
t_0 x_1t_1\cdots t_{n-1}x_n=t'_0x'_1\cdots t'_{p-1}x'_p. 
\end{equation}
Without loss of generality, we assume that $|t_0|\le|t'_0|$;
let $k$ be the greatest non-negative integer such that a word $s$ exists with  $t_0 x_1t_1\cdots x_{k}t_{k}s=t'_0$, with $s\in A^*$.
By contradiction, we assume $s\neq\varepsilon$.
Let $z_0\in Z$ ($z_1\in Z$) be the unique word such that $t'_0\in A^*z_0$ ($t_k\in A^*z_1$). 
According to the preceding property (\ref{Eq3}), an integer  $i\in{\mathbb Z}$  and two letters $b$, $\overline{b}$ exist such that 
$z_0=\overline {b}^{|y|}\theta^i(y)b^{|y|}$. Moreover, since we have $y\not\in F(X^*)$, and since $X$ is $\theta$-invariant, we have $\theta^i(y)\not\in F(X^*)$.
By construction, the set with elements $t_0x_1\cdots t_kx_{k+1}$ and $t'_0$ is not prefix; more precisely, exactly one of the two following main conditions holds:
\begin{enumerate}
\item
At first, we  assume that $t'_0\in P(t_0x_1\cdots t_kx_{k+1})$ that is, $s\in P(x_{k+1})$ (cf. Figure 3). 
By construction, at least one of the two words $s,z_0$ is  a suffix of the other one.
Actually, since we have $z_0\not\in F(X^*)$, necessarily $s$ is  a proper suffix of $z_0$, therefore $(z_1$, $z_0)$ is an overlapping pair of words. 
According to Lemma \ref{overlapZ},
necessarily we have $|s|\ge 2|y|$, which implies $\theta^i(y)\in F(s)$: a contradiction with $\theta^i(y)\not\in F(X^*)$.
\item
Now, we assume that  $t_0x_1\cdots t_kx_{k+1}$ is a proper prefix of $t'_0$, thus $x_{k+1}\in P(s)$: we have $s=x_{k+1}s_1$, with $s_1\neq\varepsilon$.
Let $z_2\in Z$ be the unique word such that  $t_{k+1}\in z_2A^*$.
By construction the set with elements $t_0x_1\cdots t_kx_{k+1}z_2$ and $t'_0$ is not prefix. More precisely, exactly one of the two following cases occurs:
\begin{figure}
\begin{center}
\includegraphics[width=12cm,height=5cm]{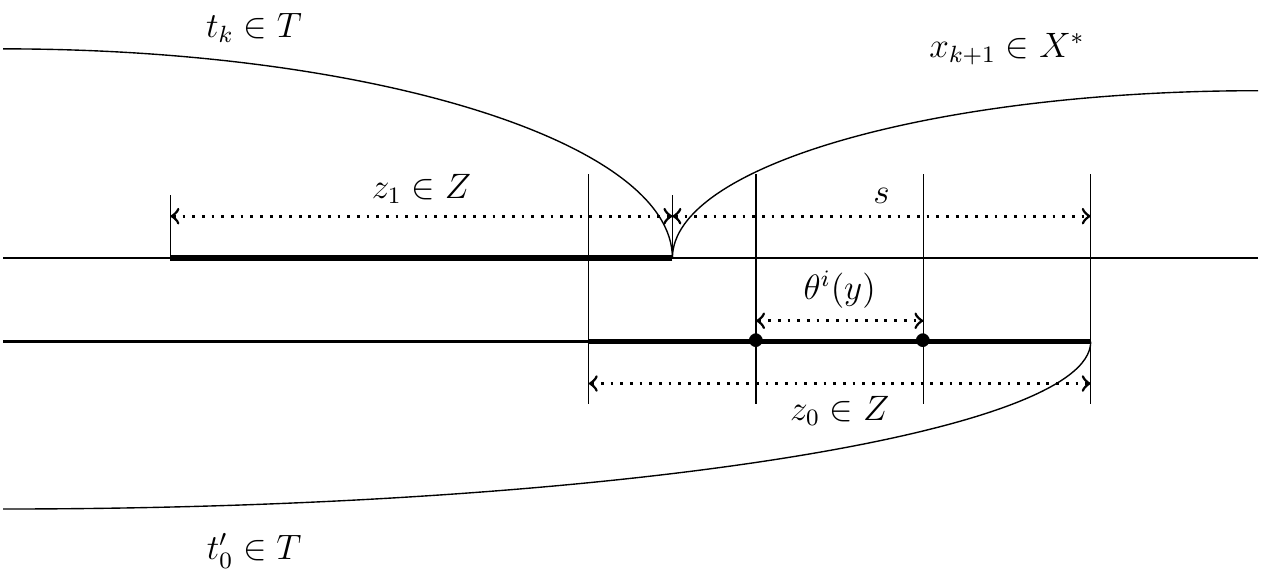}
\end{center}
\caption[]
{Proof of  Lemma  \ref{TcupXcode} - the case where $s\in P(x_{k+1})$.}
\end{figure}
\begin{enumerate}[2.1]
\item
The first case corresponds to $t'_0$ being a proper prefix of $t_0x_1\cdots t_kx_{k+1}z_2$, that is 
$s_1$ being  a proper prefix of $z_2$ (cf. Figure 4).
With this condition,   the word $z_0$ is necessarily a factor of $z_1x_{k+1}z_2$.
According to Lemma \ref{facteurInterne},
since we have $s_1\neq\varepsilon$, this implies $z_1=z_0$, which contradicts $s\neq\varepsilon$.
\begin{figure}
\begin{center}
\includegraphics[width=12cm,height=5cm]{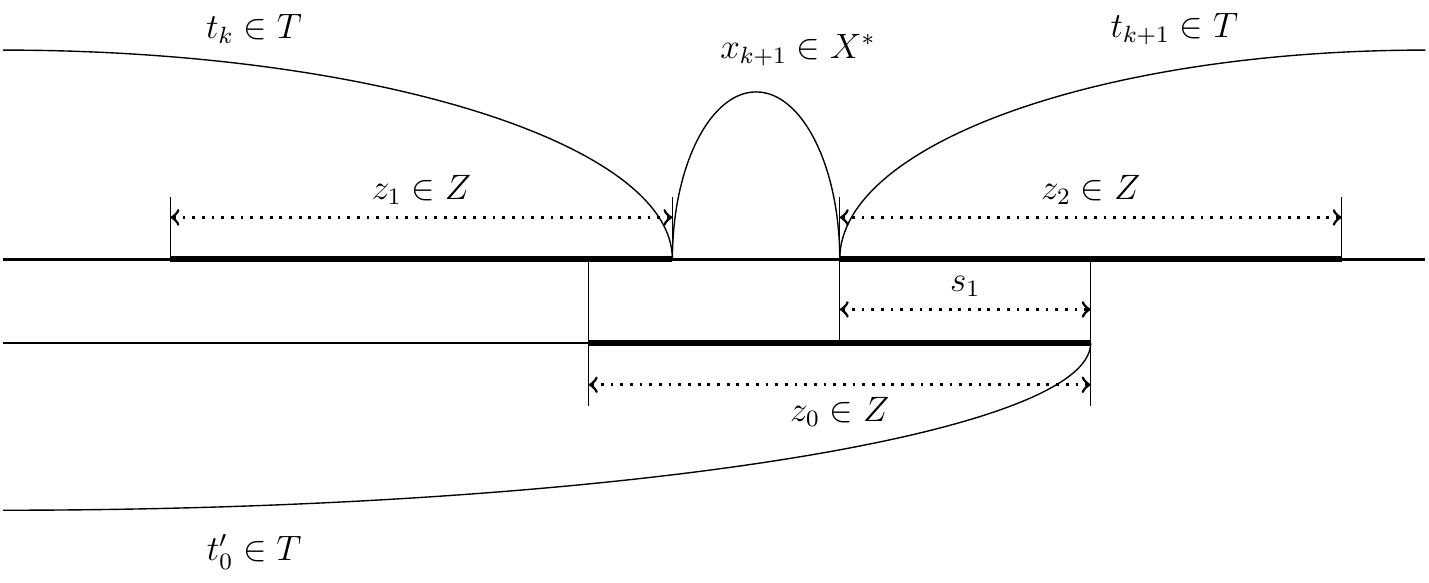}
\end{center}
\caption[]
{Proof of  Lemma  \ref{TcupXcode} -the case where $x_{k+1}\in P(s)$ and $s_1\in P(z_2)\setminus\{z_2\}$.}
\end{figure}
\item
It remains to consider the case where $z_2$ is a  prefix of $s_1$ (cf. Figure 5). 
Actually, since $t_0x_1\cdots t_kx_{k+1}z_2$ is  a prefix of $t_0x_1\cdots t_kx_{k+1}s_1=t'_0$, 
with $|z_2|=|z_0|$, necessarily $z_0$ is  a suffix of $s_1$, hence we have $s_1\in z_2A^*\cap A^*z_0$, thus $s_1\in W$ according to (\ref{W}).
We obtain $t'_0=t_0x_1\cdots t_k x_k s_1\in (TX^*)^+W$, thus $t'_0\in (T\cup X)^+W$:
this contradicts  (\ref{T}).
\begin{figure}
\begin{center}
\includegraphics[width=12cm,height=5cm]{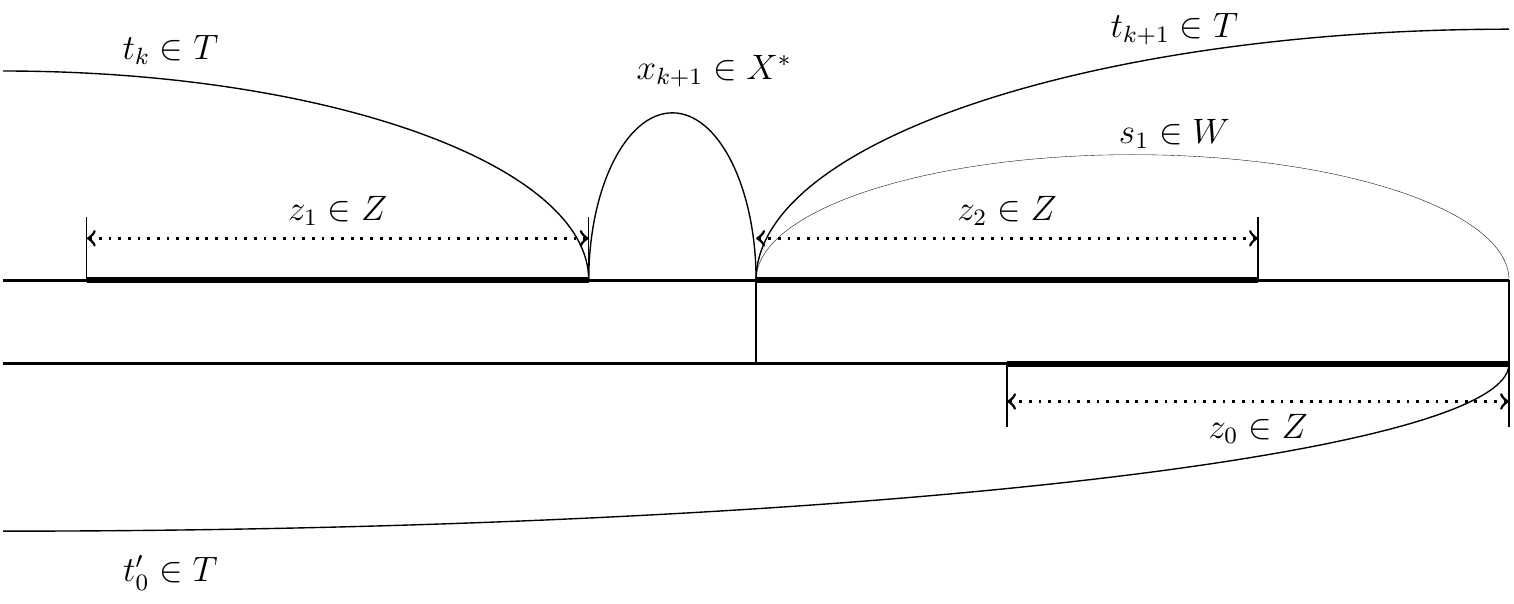}
\end{center}
\caption[]
{The case where $x_{k+1}\in P(S)$, with $z_2\in P(s_1)$ in the proof of  Lemma  \ref{TcupXcode}.}
\end{figure}
\end{enumerate}
\end{enumerate}
In each case we obtain a contradiction: as a consequence we have $s=\varepsilon$, thus $t'_0=t_0 x_1 \cdots x_{k}t_{k}$. 
Once more according to (\ref{T}), it follows from $t'_0\in T$ that we have $k=0$, thus $t'_0=t_0$.  As a consequence, Equation (\ref{equation1}) is equivalent to the following one:
\begin{equation}
\label{equation2}
x_1t_1\cdots t_{n-1}x_n=x'_1\cdots t'_{p-1}x_p. 
\end{equation}
By iterating these arguments, we shall obtain:
$n=p$ and $x_i=x'_i$, $t'_j=t_j$ $(0\le i\le n$, $0\le j\le n-1$), therefore $X\cup T$ is a code:
this completes the proof of Lemma \ref{TcupXcode}.\cqfd
\end{pot1}
\begin{lem}
\label{complete}
The code $X\cup T$ is complete.
\end{lem}
\newproof{pot2}
{\bf Proof}
\begin{pot2}
Let $w\in A^*$. According to the construction of $W$ we have $ZwZ\subseteq W$. 
According to  (\ref{decomposable}) this implies $ZwZ\subseteq (X\cup T)^*$, therefore we have $w\in F((X\cup T) ^*)$.\cqfd
\end{pot2}
In the case where $A$ is a finite alphabet, the (anti-)automorphism $\theta$ is of finite order. If $X$ is a regular code, in starting with $y\not\in F(X^*)$, the construction in (\ref{Z}) leads to a finite set $Z$: this guarantees the regularity of the sets $W$ and $T$.
As a direct consequence, we obtain the following result:
\begin{theorem}
\label{completion} 
Given a non-complete $\theta$-invariant code  $X\subseteq A^*$, the two following properties hold:

(i) In any case, $X$ can be embedded into a complete $\theta$-invariant code in $A^*$.

(ii) If $A$ is finite and $X$ regular, then $X$ can be embedded into a regular complete $\theta$-invariant code in $A^*$.
\end{theorem}
\begin{expl}
\label{E4}
Let $A=\{a,b\}$, and $\theta$ be the anti-automorphism such that $\theta(a)=b$, $\theta(b)=a$, and let $X=\{a^4, a^2b^2, a^2b^4, a^4b^2, ba, ba^4, b^4a, b^4\}$\\
Trivially, $X$ is $\theta$-invariant. By applying Sardinas-Patterson algorithm \cite[§    2.3]{BPR10}, one can easily verify that $X$ is  a (non-prefix) code.
It is non complete: by making use of the uniform Bernoulli distribution $\pi$, we obtain $\pi(X)<1$.\\
Let $y=ba^3ba$. 
Firstly, we note that we have $y\not\in F(X)$, hence $y\in F(X^*)$ implies $y=sp$, with $s\in S(X)$ and $p\in P(X^*)$.
Secondly, we have $S(X)\cap P(y)=\{b,ba\}$,
but since $\{a^3ba, a^2ba\}\cap P(X^*)=\emptyset$, necessarily we have $y\not\in F(X^*)$.
Thirdly, in view of obtaining an overlapping-free word, we substitute $by=b^2a^3ba$ to $y$ (we have $by\not\in F(X^*)$. \\
With the notation (\ref{Eq2}, \ref{Z}), we have $z={a^{7}} b^2a^3bab^{7}$, thus:\\ $Z=\bigcup_{i\in {\mathbb Z}}\{\theta^i(z)\}=\{a^{7} b^2a^3bab^{7}, a^7bab^3a^2b^7\}$.
Moreover, the sets $W$ and $T$ shall be constructed according to (\ref{W},\ref{T}).
\end{expl}
Example \ref{E4} provides a (non-finite) regular complete $\theta$-invariant code;
 in the sequel we give an example of a non-regular one:
\begin{expl}
\label{E5}
Let $A=\{a,b\}$,  and $\theta$ be an arbitrary (anti-)automorphism of $A^*$. Consider the  famous Dyck language $D_1^*=\{w\in A^*:|w|_a=|w|_b\}$.
Each of its elements is classically represented by a so-called Dyck path in the grid ${\mathbb N}\times {\mathbb Z}$.
To be more precise,  with each word $w=w_1\cdots w_n$ (with $w_i\in A$, for $1\le i\le n$), a unique  path  is associated, namely $(i,y_i)_{0\le i\le n}$, with $y_0=y_n=0$ and such that, for each $i\in [1,n]$: 
$$w_{i}=a\Longrightarrow y_{i}=y_{i-1}+1~~~~{\rm and}~~~~  w_{i}=b\Longrightarrow y_{i}=y_{i-1}-1.$$
By construction,  $D_1^*$ is  a free submonoid of $A^*$. Its  minimal generating set  is the so-called Dyck code $D_1$, whose elements are represented by those of the preceding non-empty paths which satisfy the following condition:
$$n\ge2~~~~{\rm  and}~~~~(\forall i\in [1,n-1])~~y_i\neq 0.$$
The code $D_1$  is well known for being a (non-thin) complete context-free language (e.g. \cite[Example 2.5.3]{BPR10}). Moreover, according to Proposition \ref{base}, since $D_1^*$ is $\theta$-invariant, the same holds to the Dyck code.
\end{expl}
As a consequence of Theorem \ref{completion}, we obtain the following result, which states a property similar to \cite[Theorem 2.5.16]{BPR10} in the framework of $\theta$-invariant code:
\begin{theorem}
\label{equivalences}
Given a thin $\theta$-invariant code  $X\subseteq A^*$,
 the following conditions are equivalent:

{\rm (i)} $X$ is complete.

{\rm (ii)} $X$ is a maximal code.

{\rm (iii)} $X$ is maximal in the family  $\theta$-invariant codes.

{\rm (iv)} A positive Bernoulli distribution $\pi$ exists such that $\pi(X)=1$.

{\rm (v)} For any positive Bernoulli distribution $\pi$, we have $\pi(X)=1$.
\end{theorem}
\newproof{pot3}{\bf Proof}
\begin{pot3}
According to \cite[Theorem 2.5.16]{BPR10}, the conditions (i), (ii) , (iv), (v) are equivalent.
Trivially, Condition (ii) implies Condition (iii). By contradiction, we prove   that Condition (iii) implies Condition (i). Starting with
a non-complete $\theta$-invariant code $X$, according to Theorem \ref{completion} 
the existence of a complete $\theta$-invariant code that strictly contains $X$ is guaranteed, thus $X$ is not maximal in the family of $\theta$-invariant codes:
this completes the proof.
\cqfd
\end{pot3}
\section*{Aknowledgement}
We would like to thank the anonymous  reviewers for their fruitful suggestions and comments.
\bibliographystyle{plain}
\bibliography{mysmallbib}{}
\end{document}